# Diffusion Quantum Monte Carlo Study of Martensitic Phase Transition: The Case of Phosphorene

Kyle G. Reeves*, Yi Yao*, and Yosuke Kanai#

Department of Chemistry, University of North Carolina at Chapel Hill, Chapel Hill, NC, USA

*Equal contributions.
# To whom correspondence should be addressed: ykanai@unc.edu

**Abstract**
Recent technical advances in dealing with finite-size errors make quantum Monte Carlo methods quite appealing for treating extended systems in electronic structure calculations, especially when commonly-used density functional theory (DFT) methods might not be satisfactory. We present a theoretical study of martensitic phase transition of a two-dimensional phosphorene by employing diffusion Monte Carlo (DMC) approach to investigate the energetics of this phase transition. The DMC calculation supports DFT prediction of having a rather diffusive barrier that is characterized by having two transition states, in addition to confirming that the so-called black and blue phases of phosphorene are essentially degenerate. At the same time, the calculation shows the importance of treating correlation energy accurately for describing the energy changes in the martensitic phase transition, as is already widely appreciated for chemical bond formation/dissociation. Building on the atomistic characterization of the phase transition process, we also discuss how mechanical strain influences the stabilities of the two phases of phosphorene.

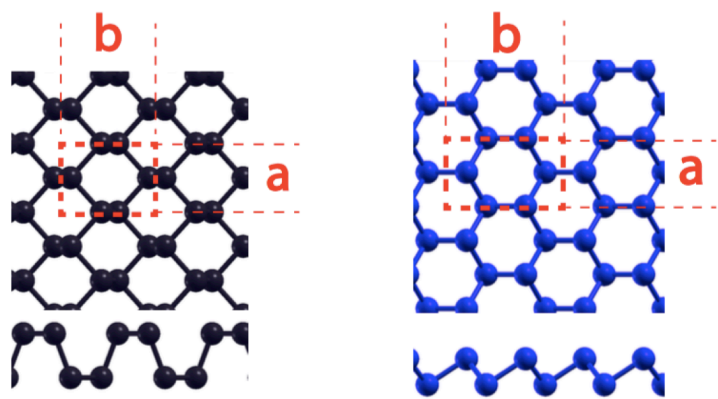

**Figure 1.** The structures of phospherene in the black (LEFT) and blue (RIGHT) phases, with top (TOP) and side (BOTTOM) views. The unit cell consists of 4 phosphorus atoms. Lattice parameters a and b are also indicated (see main text).

## Introduction

Phosphorene is the two-dimensional counterpart of layered black phosphorus, and a detailed investigation of this particular material is of great interest for various technological applications because the material is predicted to exhibit unique electronic and optical properties(1, 2). The most commonly discussed phase is black phosphorene, and the material has already been synthesized and has shown good performance in a few applications for solar energy conversion(3, 4). In additional to the black phosphorene phase, more than ten different phases of phosphorene have been predicted to exist in literature (1, 2, 5). Among these different phases of phosphorene, the so-called blue phosphorene phase is predicted by density functional theory (DFT) to have a similar honeycomb structure and ground state energy as the black phosphorene phase yet possessing distinctively different electronic properties(2). Despite these theoretical predictions, however, it has yet to be experimentally synthesized. In this work, we investigate the possibility of martensitic phase transition from the common black phosphorene phase to blue phosphorene phase by employing diffusion quantum Monte Carlo calculation(6) to obtain accurate reaction energetics along the transition pathway.

Several works have explored the possibility of diffusionless phase transitions between different phases because such phase transition can be facilitated through an application of external strain on the material(2). Determination of the relevant reaction coordinate and obtaining accurate reaction energetics along the reaction coordinate are two important requirements in theoretical investigations of such transitions. Generally, minimum energy path (MEP) defines the reaction coordinate and transition states of infrequent transition events such as chemical reactions when the entropic contribution is not dominant, and various numerical methods have been proposed over the years for locating the MEP(7) include chain-of-states methods such as nudged elastic band (NEB) method(8) and the string method(9) as well as other types of numerical approaches, which are based on local curvatures of the potential energy surface such as the dimer method(10) and the activation-relaxation technique(11) for locating transition states. Unlike for chemical reactions, however, phase transitions involve not only changes of atom coordinates but also changes in the lattice vectors. Thus, the lattice degrees of freedom need to be included in the search space for locating the MEP. For the particular case of the phosphorene phase transition in this work, we are concerned with a martensitic (diffusionless) phase transformation in which the phase changes involve cooperative/concerted movement of all atoms without involving long-range diffusion of atoms. In this work we employ the so-called variable cell NEB method(12) to locate the MEP in the extended configuration space, which includes lattice coordinates so that the reaction coordinate can be defined even for the phase transition.

Another important aspect of the present study is the electronic structure calculation that is used to obtain reaction energetics for large extended systems. For theoretical investigations of extended systems, density functional theory (DFT) calculations have become quite common because of their balanced accuracy and computational affordability. However, while geometries appear to be well described, reaction energetics are often poorly described by DFT calculations when employed with most common exchange-correlation (XC) approximations. In this regard, a higher-level of first-principles approaches is desired for accurately characterizing reaction energetics. Among many-body first-principles approaches, diffusion quantum Monte Carlo (DMC) method is promising. Indeed, its accuracy often rivals that of CCSD(T) method with a large basis set even with the fixed node approximation using only one Slater determinant when the systems do not show strong correlation behaviors (e.g. no chemical bond formation/dissociation) (13-16). At the same time, application of explicitly many-body methods such as DMC calculation to extended systems like phosphorene is not straightforward because of the many-body finite-size error that is associated with using the periodic boundary conditions to describe infinitely large systems (17). There are already extensive literatures on how the many-body finite-site error can be corrected in the context of DMC(18-22), and the topic remains an important area of further pursuit. The finite-size error in DMC is very different from the finite-size error in DFT, which is entirely due to a finite sampling of Brillouin zone. Detailed discussion of the origin of the finite-size error in DMC can be found, for example, in Ref. (22).our study on the martensitic phase transition of the extended phosphorene, we examine both the extrapolation approach (23) as well as the DFT-based correction scheme by Kwee et al(24), along with employing the twist-averaging approach for Brillouin zone integration(25).

**Theoretical Methods**

*Phase Transition Pathway:*
The variable cell nudged elastic band (VCNEB) method is used to locate the minimum energy path (MEP) for the phase transition between blue and black phosphorene. VCNEB is a nudged elastic band (NEB) method that allows unit cell vectors to be varied in addition to the atomic positions. (12) This builds on the NEB framework by including unit cell vectors in the configuration space during the search for the MEP. The configuration space can be expressed as

$$X = (\varepsilon_{1i}, \varepsilon_{2i}, \varepsilon_{3i}; r_1, r_2, \ldots, r_N) \text{ (i = 1,2,3)}$$

where $\varepsilon_{ij}$ is the elements in the finite strain tensor $\bar{\bar{\varepsilon}}$ which is defined through $h = (1 + \bar{\bar{\varepsilon}})h_0$. $h$ is the matrix of lattice vectors and $h_0$ is a reference matrix of lattice vectors, which was taken to be that of the blue phosphorene. In the VCNEB calculation, the configuration space is represented by a total of $9 + 3N_{atom}$ variables. The convergence threshold for the VCNEB calculation was set to 0.005 eV/Å, and we use a total of 47 images to represent the path between the initial and final states for the phase transition. We performed the VCNEB calculation by interfacing the USPEX code (26) with Quantum Espresso code (27) such that the underlying electronic structure is derived from DFT calculation. The PBE exchange-correlation (XC) functional(28) was employed, and Vanderbilt ultrasoft pseudopotentials(29) were used for describing core-valence electron interactions. The planewave cutoff for Kohn-Sham wavefunctions and the smooth charge density cutoff were 20Ry and 200Ry, respectively. The Monkhorst-Pack k-point grid of 8x8x1 was used for sampling the Brillouin Zone using the unit cell with four atoms. For stationary states along the located MEP, further DFT calculations were performed with PBE0 hybrid XC functional for additional comparison. In this case, Troullier-Martin norm-conserving pseudopontials(30) were used with the planewave cutoff for Kohn-Sham wavefunctions of 80 Ryd.

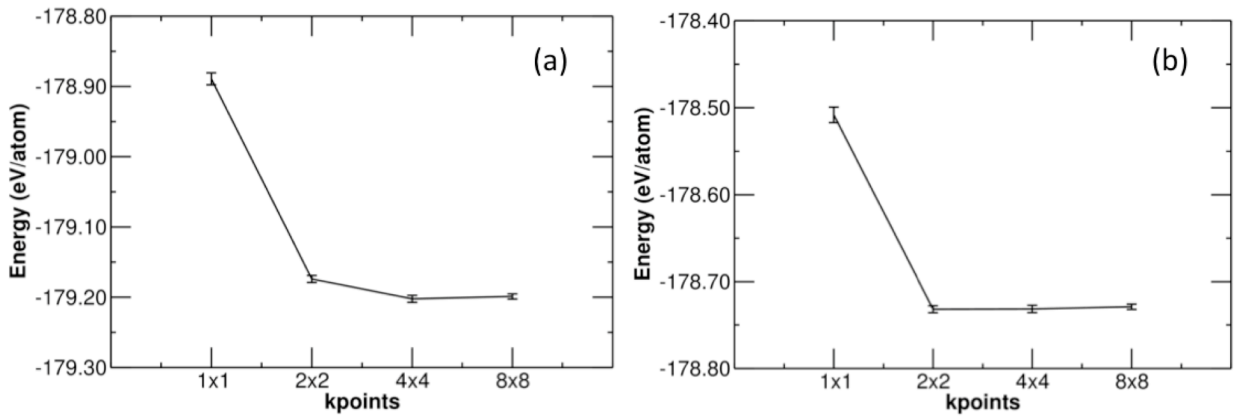

**Figure 2**. The convergence of BZ sampling in the twist averaging for DMC calculations for (a) blue phospherene phase and (b) TS2 structure (see Figure 4) using the 2x2 supercell (64 atoms). The calculations with 2x2 k-points and 8x8 k-points differ in the total energy by less than 0.01 eV/atom.

*Reaction Energetics:*
Fixed-node diffusional Monte Carlo (DMC) calculations were performed using the QWALK code(31) to obtain accurate reaction energetics at the stationary points along the MEP located. Fermion nodes here were given by Slater determinant of Kohn-Sham orbitals, which were obtained with PBE functional using the CRYSTAL code(32). $3s^2\ 3p^3$ electrons of phosphorous atoms are treated explicitly as valence electrons using the BFD pseudopotential(33) where the Ne core is treated within the pseudopotential. For the basis sets, we use a modified BFD basis with the modification scheme suggested by Lucas Wagner. The basis set in the format of CRYSTAL14 code is provided in Supporting Information. Locality approximation is used for dealing with the non-local part of the pseudopotential(33). Importance sampling was introduced with trial wave functions, which were given by the Slater determinant multiplied by one-body and two-body Jastrow correlation factor. Variational QMC calculation was used with the variance minimization(34) to obtain the trial wave functions. As implemented in the QWALK code, the DMC algorithm by Umrigar, Nightingale, and Runge(35) was used with a modification to the branching part for keeping the number of walkers constant(31). We used the imaginary time step of 0.01 a.u. to perform the DMC calculations, and the convergence test is presented in Supporting Information. In recent 2016 work, Zen *et al*. has shown that DMC energies could suffer the size consistency when the time step is not sufficiently small(36). The authors proposed an alternative DMC algorithm that significantly reduces the size-consistent problem, allowing the use of larger time steps. We, however, did not implement this new algorithm in this work but instead use the rather small time step (0.01 a.u.) with the above-mentioned modified UNR algorithm.

*Finite-size Error Corrections in Quantum Monte Carlo:*
Independent-particle methods such as DFT can employ the Bloch theorem to eliminate the finite-size error by obtaining the macroscopic limit via Brillouin zone (BZ) integration. Finite-size error is one of the major sources of error in quantum Monte Carlo calculation of extended systems(6). There are two sources contributing to the finite-size error: one-body and two-body errors. The one-body contribution to the error is related to BZ sampling and can be largely removed using twist averaging, an approach discussed in Ref (25). We used 4 k-points (2x2x1) with a 64-atom simulation cell for the "twist". We show the convergence test in Figure 2 for the blue phase and also for an intermediate transition state structure along the MEP because the latter was found to be metallic.

The two-body error is more problematic, and it originates from the artificial periodicity in the long-range electron-electron Coulomb interaction when extended systems are simulated using periodic boundary conditions. The most straight-forward approach (and most rigorous, in principle if there are no size-consistency errors associated with the DMC algorithm as discussed in above) is an extrapolation using increasingly larger simulation cells, as was utilized in a recent work for a 2D material by Wagner and co-workers (23). Assuming that phosphoeres can be considered as a two-dimensional system, one expects the total energy to scale as $N^{-5/4}$ where N is the system size according to the work by Drummond et al(22). Fitting DMC energies as a linear function of $N^{-5/4}$ where N is the number of unit cells in the supercell, we can extrapolate the linear fit to the limit of $N^{-5/4}=0$ (i.e. an infinitely large simulation cell).

Alternatively, Kwee *et al.* proposed a scheme that is based on DFT calculations using a modified finite-size local density approximation (LDA) to XC functional for estimating the two-body error (24), and it has thus far has been applied successfully to bulk systems (37, 38). A computational advantage of this approach is that finite-size correction can be approximated using DFT calculations rather than performing a series of more expensive QMC calculations for the extrapolation. Whereas the LDA XC functional was originally parameterized using QMC results for an infinite jellium (39), this finite-size LDA XC functional is instead parameterized to fit finite size QMC calculations of jellium, thus the finite size error is intentionally retained in the modified XC functional itself. The Kohn-Sham single-particle Hamiltonian can be expressed as

$$H_{DFT} = -\nabla^2 + V_{ion} + V_H(\boldsymbol{r}) + V_{xc}^\infty(\boldsymbol{r})$$

where the first three terms are kinetic energy, ionic potential energy, and Hartree term respectively, and the last term is the XC potential. This term is calculated as $V_{xc}^\infty(\boldsymbol{r}) = \delta(n(\boldsymbol{r})\epsilon_{xc}^\infty(n))/\delta n(\boldsymbol{r})$ where $n(\boldsymbol{r})$ is the charge density and $\epsilon_{xc}^\infty(n)$ is obtained from the QMC calculation of the infinite size jellium. In the finite-size error LDA functional by Kwee *et al.*, the XC energy has the form of $\epsilon_{xc}^{FS}(n) \equiv \epsilon_x(r_s, L) + \epsilon_c(r_s, L)$, where $r_s$ specifies the density of the jellium via $\frac{4\pi r_s^3}{3} \equiv \frac{1}{n}$ and L denotes the linear size of the supercell (for example in cubic cell, $L = V^{1/3}$). The finite size error then can be estimated using DFT calculations as

$$\Delta DFT^{FS} = E(\infty) - E^{FS}(L)$$

where the $E(\infty)$ is the DFT total energy (which is converged with respect to k-point integration) using the original LDA XC functional and $E^{FS}(L)$ is the total energy using the modified finite-size LDA XC functional with the same k points as in the QMC calculation. This correction can then be used to correct the QMC calculated value to estimate the QMC energy in the limit of an infinite cell as

$$E_{QMC}^\infty = E_{QMC}^{FS} + \Delta DFT^{FS}$$

where $E_{QMC}^{FS}$ is the finite-size QMC result and $E_{QMC}^\infty$ is QMC result in the limit of infinite size. This approach has been employed successfully for studying materials with a simple cubic cell in literature(24). In our work calculations, we treat phosphorenes as strictly a two-dimensional system, and we take the linear size L to be the square root of the area of the plane of the simulation cell that is parallel to the phosphorene (xy plane). We checked the dependence of the $E_{QMC}^\infty$ on the size of the vacuum region along z direction and found it to be negligibly small (less than 0.0005 eV/atom).

We compared the above-mentioned extrapolation approach and the correction scheme using the modified finite-size XC functional within DFT calculations (KZK correction). Supercells with 2x2, 3x3, and 4x4 cell sizes in blue phosphorene phase were considered here, and they correspond to N=4, 9, and 16 of the $N^{-5/4}$ extrapolation. Figure 2 shows that QMC total energies with and without the KZK correction as a function of $N^{-5/4}$. The extrapolation with the uncorrected total energies using the least-square fitting gives -178.9190 eV/atom in the limit of $N^{-5/4}=0$. The KZK-corrected total energy gives a similar value -178.9119 eV/atom when the simulation cell size with N=4x4=16 is used, resulting in the difference of only 0.0071 eV/atom. Given the very small difference, which is certainly below the necessary accuracy for our work, we chose the computationally less demanding correction scheme using the KZK functional for calculating $E_{QMC}^{\infty}$ using the simulation cell of N=4x4=16 for the rest of the paper.

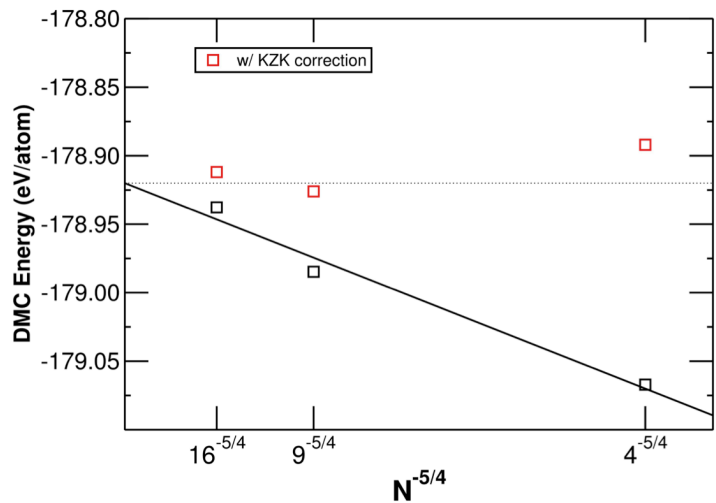

**Figure 3.** DMC total energies with and without the KZK correction for the blue phosphorene phase using 2x2 ($N^{-5/4}= 4^{-5/4}$), 3x3 ($N^{-5/4}= 9^{-5/4}$), and 4x4 ($N^{-5/4}= 16^{-5/4}$) supercells. The dashed black line indicates the extrapolated value to the thermodynamic limit using the DMC total energies without the KZK correction (solid black line). The standard deviations on the DMC total energies are all smaller than the size of the square symbols.

**Results and Discussion**
*Minimum Energy Path:*
Previously, Zhu *et al.* estimated an upper bound of the transition energy barrier to be approximately 0.47eV/atom using DFT calculation with PBE XC functional. The calculation assumed the out-of-plane displacement of one atom to be the reaction coordinate for the phase transition(2). Here, we employed the variable cell nudged elastic band (VCNEB) method to locate the minimum energy path (MEP) between black and blue phosphorene phases as shown in Figure 4. The transition involves significant changes to atom positions as well as to lattice vectors as shown in Figure 5. The transition from black phosphorene to the first transition state (TS1) has the energy change of 0.308 eV/atom, according to DFT-PBE calculation. The cell parameter *a* changes from 3.30Å to 3.48Å; while, the cell parameter *b* increases from 4.58Å to 5.15Å. The local minimum (LM) is only 0.024 eV/atom lower in energy compared to TS1 and 0.052 eV/atom lower in energy compared to the second transition state (TS2). These values indicate that the local minimum is highly unstable even at room temperature. The cell parameters in the region of the LM vary significantly while the total energy remain essential unchanged suggesting a structure with a small Young's Modulus for the LM structure. From TS1 to TS2, the cell parameter *a* (see Figure 5) decreases slightly from 3.48Å to 3.27Å, and the cell parameter *b* increases significantly from 5.15 Å to 5.90Å. The potential energy decreases by 0.336 eV/atom from TS2 to the blue phosphorene phase. In this part of the transition the lattice parameters remain almost the same changing by only 5% from the cell parameters of TS2 while significant atomic displacements accompany the transition. DFT-PBE calculation yields the overall energy barrier of 0.336 eV/atom for the phase transition.

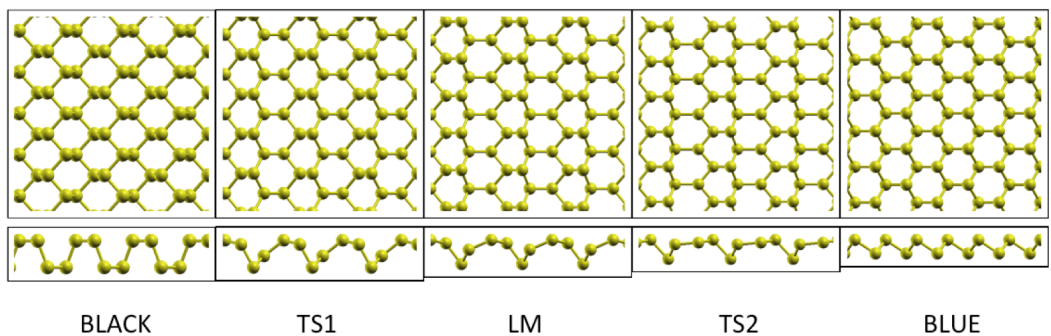

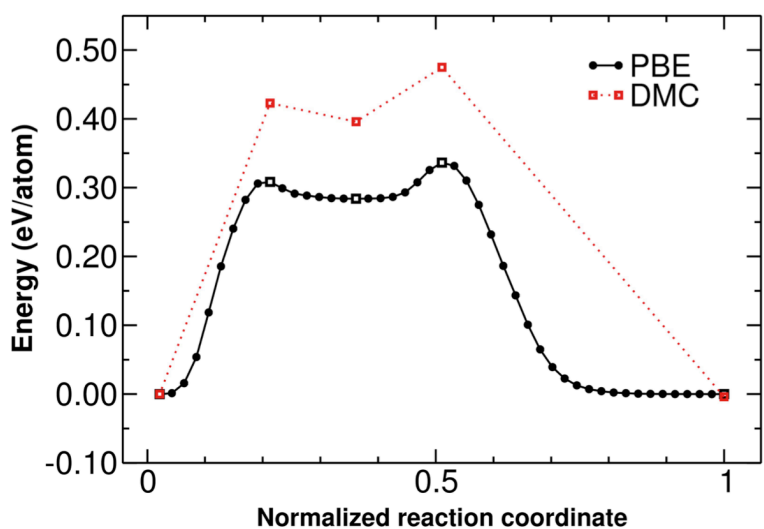

**Figure 4.** (TOP) Structures of the stationary points along the minimum energy path (MEP) for the phase transition between black phosphorene and blue phosphorene as calculated using variable cell nudged elastic band method based on DFT-PBE forces. Black phosphorene, transition state 1 (TS1), local minimum (LM), and transition state 2 (TS2), and blue phosphorene are shown. (BOTTOM) The energetics along MEP according to the DFT-PBE calculations (black line), and DMC calculations for the five stationary states (red boxes).

*Reaction Energetics:*

We performed DMC calculations at the five stationary structures along the MEP (reaction coordinate). In particular, the existence of the metastable structure needs to be verified. In DMC calculations of periodic systems, a large error may come from the finite size effects. We utilized two approaches to calculate and to reduce the finite size error as discussed in Theoretical Methods section. Consistent with the DFT result, DMC calculation confirms that black and blue phosphorene phases have almost the same total energies where black and blue phosphorene structures yield -179.0611±0.0016 eV/atom and -179.0647±0.0015 eV/atom, respectively. Importantly, the metastable LM structure is located energetically lower than the two TS structures accordingly to DMC calculation as well, confirming the existence of this metastable local minimum. However, DMC predicts a slightly greater energy barrier for over all transition than DFT. The DMC calculation yields a barrier of 0.4237±0.0027 eV/atom for the transition from blue phosphorene to TS1, which is about 1.4 times the value predicted by DFT (0.308 eV/atom). The energy difference between TS1 and LM is 0.0283±0.0024 eV/atom compared to 0.024 eV/atom in DFT. The energy difference between TS2 and LM is 0.0793± 0.0029 eV/atom compared to the DFT result of 0.052 eV/atom. The energy change from TS2 to the black phosphorene phase (0.4783±0.0030 eV/atom) is again about 1.4 times the DFT value (0.336 eV/atom).

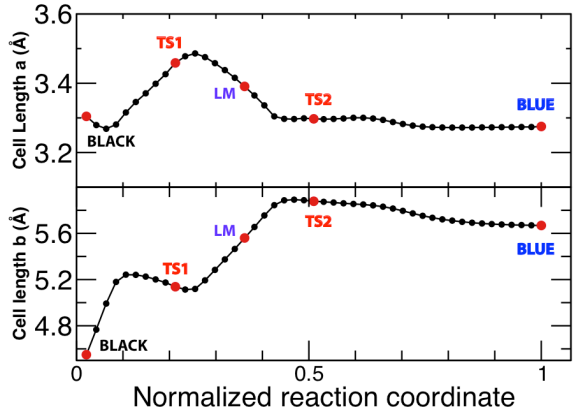

**Figure 5.** The variation of the cell parameters along the minimum energy path (MEP) as calculated using the variable cell nudged elastic band method based on DFT-PBE forces. Cell parameters a and b (See Figure 1) are shown in the top and bottom plots respectively. Stationary states along the MEP are indicated with red circles.

*Comparison of DFT-PBE0 and DMC calculations:*

In recent years, PBE0 XC approximation(40) has become quite popular for investigating extended systems due to its superior performance compared to standard GGA approximations such as PBE. In terms of computational cost, PBE0 is approximately one-to-two orders of magnitude more expensive than PBE using planewaves as basis set but still much more affordable than performing DMC calculations. A number of studies on molecular systems have shown that the GGA functionals with some amount of exact exchange describe the reaction barriers significantly better as compared to PBE(41, 42). In the present case of phosphorene martensitic phase transition, there is also a sizable improvement with PBE0 in the direction toward DMC calculation result as shown in Table I. For both TS1 and TS2 energy barriers, PBE0 calculation predicts the values that are in-between DFT-PBE and DMC calculations. We also observe the relative energy of the local minimum is improved while the overall energy difference between the black and blue phases is essentially unchanged.

*Effects of External Strain:*

Most experimental studies on phosphorenes are focused on the black phosphorene phase(3, 4) while we confirmed here the earlier DFT prediction that the blue phosphorene phase is equally stable energetically by employing accurate quantum Monte Carlo calculations. Having shown that DFT calculations can be trusted for the qualitative description of the Martensitic phase transition, we study the extent to which an application of external strain can lower the transition energetics. Since the lattice vectors of these two phases differ significantly along the b-vector direction (see Figure 5), we examined whether an application of external strain might induce a more favorable transition energetics for the blue phosphorene. In their DFT work, Hu and Dong have previously studied the effect of applied strain and showed that the blue phosphorene phase can be made energetically more stable than the black phase (43). Such an application of strain is experimentally conceivable as demonstrated for graphenes and nanowires as they are suspended over a trench(44). We mimic such an experimental setting, by constraining the b-vector of all 47 images in the VCNEB calculation to be 100%, 110%, and 120% of the blue phosphorene so that we could examine how the MEP as well as the reaction energetics change. The structures along all the MEPs do not change substantially, except that a new local minimum appears between the black phosphorene and TS1 structures for the 120% case as shown in **Figure** . In all cases under the strain, the blue phosphorene phase is now approximately 0.1 eV/atom lower in energy than the black phosphorene phase. At the same time, the energy barrier (in DFT-PBE calculations) changes substantially from 0.34 eV/atom to 0.25 eV/atom to 0.12 eV/atom to 0.05 eV/atom as the b-vector is constrained to larger values.

**Table I.** Relative energetics for each stationary state (see Figure 3) along the minimum energy path (MEP) with respect to the black phosphorene. The table provides a comparison between DFT calculations using PBE and PBE0 as well as to diffusion Monet Carlo (DMC) calculations. All units are in eV/atom and the statistical uncertainty is indicated by parentheses for the DMC calculations.

| Method | Black | TS1 | LM | TS2 | Blue |
|---|---|---|---|---|---|
| PBE | 0.000 | 0.307 | 0.282 | 0.335 | -0.001 |
| PBE0 | 0.000 | 0.357 | 0.347 | 0.397 | 0.004 |
| DMC | 0.000 | 0.423(3) | 0.396(3) | 0.475(3) | -0.004(3) |

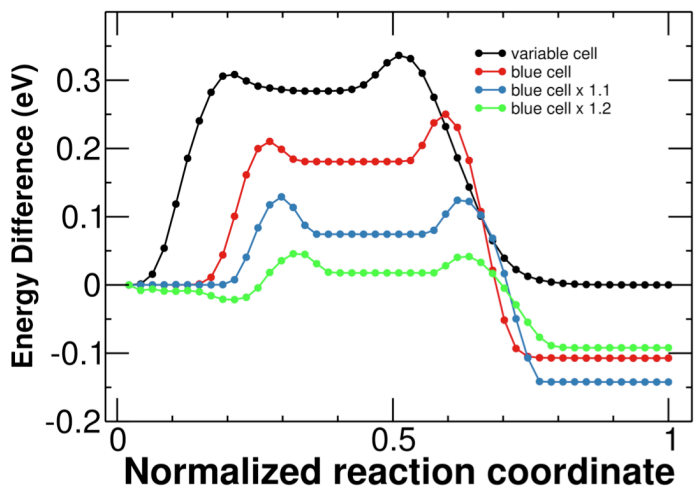

**Figure 6.** The stress-dependence of the minimum energy path (MEP) is calculated using the variable cell nudged elastic band method while simultaneously constraining the b lattice parameter. The b lattice parameter was constrained to be 100% (red), 110% (blue), and 120% (green) the length of the b lattice parameter for the fully relaxed blue phosphorene structure. The variable cell MEP (black) was included for comparison. MEPs are plotted relative to the energy of the black phosphorene structure.

**Conclusion**

In this work, we presented a theoretical study of martensitic phase transition of a two-dimensional phosphorene between its so-called black and blue phases. The martensitic phase transition pathway was first determined using the variable-cell nudged elastic band method based on the atomic forces and stress tensors from DFT calculations. DMC method was then used to investigate the energetics of the phase transition, which was found at DFT level of theory to have a rather diffusive barrier(45) characterized by two transitions states. The DMC calculations support this interesting finding in addition to confirming that the two phases are essentially degenerate. The phase transition was found to consist of two distinct stages separated such that the initial atomic displacements are accompanied by a lattice deformation, while the reorganization occurring in the second stage is nearly exclusively due to the atomic displacements alone. Having demonstrated that DFT calculations predict the qualitatively correct trend of the martensitic phase transition energetics, we also showed that application of mechanical strain considerably alters the stability of the two phases of phosphorene, and the blue phase can be made much more stable than the black phase, with accompanied lowering of the energy barrier for the phase transition.

**Acknowledgment**

We thank Lucas K. Wagner (UIUC) for helpful discussions related to quantum Monte Carlo calculations. This material is based upon work supported by the National Science Foundation under Grant No. DGE-1144081. We thank Lawrence Livermore National Laboratory for computational resources.

**Reference**

1. Guan J, Zhu Z, & Tománek D (2014) Phase Coexistence and Metal-Insulator Transition in Few-Layer Phosphorene: A Computational Study. *Physical Review Letters* 113(4):046804.
2. Zhu Z & Tománek D (2014) Semiconducting Layered Blue Phosphorus: A Computational Study. *Physical Review Letters* 112(17):176802.
3. Woomer AH, *et al.* (2015) Phosphorene: Synthesis, Scale-Up, and Quantitative Optical Spectroscopy. *ACS Nano* 9(9):8869-8884.
4. Hu J, *et al.* (2016) Band Gap Engineering in a 2D Material for Solar-to-Chemical Energy Conversion. *Nano Letters* 16(1):74-79.
5. Wu M, Fu H, Zhou L, Yao K, & Zeng XC (2015) Nine New Phosphorene Polymorphs with Non-Honeycomb Structures: A Much Extended Family. *Nano Letters* 15(5):3557-3562.
6. Foulkes WMC, Mitas L, Needs RJ, & Rajagopal G (2001) Quantum Monte Carlo simulations of solids. *Reviews of Modern Physics* 73(1):33-83.
7. E W & Vanden-Eijnden E (2010) Transition-Path Theory and Path-Finding Algorithms for the Study of Rare Events. *Annual Review of Physical Chemistry* 61(1):391-420.
8. Henkelman G, Uberuaga BP, & Jónsson H (2000) A climbing image nudged elastic band method for finding saddle points and minimum energy paths. *The Journal of Chemical Physics* 113(22):9901-9904.
9. E W, Ren W, & Vanden-Eijnden E (2002) String method for the study of rare events. *Physical Review B* 66(5):052301.
10. Henkelman G & Jónsson H (1999) A dimer method for finding saddle points on high dimensional potential surfaces using only first derivatives. *The Journal of Chemical Physics* 111(15):7010-7022.
11. Mousseau N & Barkema GT (1998) Traveling through potential energy landscapes of disordered materials: The activation-relaxation technique. *Physical Review E* 57(2):2419-2424.
12. Qian G-R, *et al.* (2013) Variable cell nudged elastic band method for studying

solid–solid structural phase transitions. *Computer Physics Communications* 184(9):2111-2118.
13. Morales MA, McMinis J, Clark BK, Kim J, & Scuseria GE (2012) Multideterminant Wave Functions in Quantum Monte Carlo. *Journal of Chemical Theory and Computation* 8(7):2181-2188.
14. Dubecký M, *et al.* (2013) Quantum Monte Carlo Methods Describe Noncovalent Interactions with Subchemical Accuracy. *Journal of Chemical Theory and Computation* 9(10):4287-4292.
15. Dubecký M, Mitas L, & Jurečka P (2016) Noncovalent Interactions by Quantum Monte Carlo. *Chemical Reviews* 116(9):5188-5215.
16. Zen A, Coccia E, Luo Y, Sorella S, & Guidoni L (2014) Static and Dynamical Correlation in Diradical Molecules by Quantum Monte Carlo Using the Jastrow Antisymmetrized Geminal Power Ansatz. *Journal of Chemical Theory and Computation* 10(3):1048-1061.
17. Azadi S & Foulkes WMC (2015) Systematic study of finite-size effects in quantum Monte Carlo calculations of real metallic systems. *The Journal of Chemical Physics* 143(10):102807.
18. Fraser LM, *et al.* (1996) Finite-size effects and Coulomb interactions in quantum Monte Carlo calculations for homogeneous systems with periodic boundary conditions. *Physical Review B* 53(4):1814-1832.
19. Williamson AJ, *et al.* (1997) Elimination of Coulomb finite-size effects in quantum many-body simulations. *Physical Review B* 55(8):R4851-R4854.
20. Kent PRC, *et al.* (1999) Finite-size errors in quantum many-body simulations of extended systems. *Physical Review B* 59(3):1917-1929.
21. Chiesa S, Ceperley DM, Martin RM, & Holzmann M (2006) Finite-Size Error in Many-Body Simulations with Long-Range Interactions. *Physical Review Letters* 97(7):076404.
22. Drummond ND, Needs RJ, Sorouri A, & Foulkes WMC (2008) Finite-size errors in continuum quantum Monte Carlo calculations. *Physical Review B* 78(12):125106.
23. Wu Y, Wagner LK, & Aluru NR (2015) The interaction between hexagonal boron nitride and water from first principles. *The Journal of Chemical Physics* 142(23):234702.
24. Kwee H, Zhang S, & Krakauer H (2008) Finite-Size Correction in Many-Body Electronic Structure Calculations. *Physical Review Letters* 100(12):126404.
25. Lin C, Zong FH, & Ceperley DM (2001) Twist-averaged boundary conditions in continuum quantum Monte Carlo algorithms. *Physical Review E* 64(1):016702.
26. Oganov AR & Glass CW (2006) Crystal structure prediction using ab initio evolutionary techniques: Principles and applications. *The Journal of Chemical Physics* 124(24):244704.
27. Paolo G, *et al.* (2009) QUANTUM ESPRESSO: a modular and open-source software project for quantum simulations of materials. *Journal of Physics: Condensed Matter* 21(39):395502.
28. Perdew JP, Burke K, & Ernzerhof M (1996) Generalized Gradient Approximation Made Simple. *Physical Review Letters* 77(18):3865-3868.
29. Vanderbilt D (1990) Soft self-consistent pseudopotentials in a generalized eigenvalue formalism. *Physical Review B* 41(11):7892-7895.
30. Troullier N & Martins JL (1991) Efficient pseudopotentials for plane-wave calculations. *Physical Review B* 43(3):1993-2006.
31. Wagner LK, Bajdich M, & Mitas L (2009) QWalk: A quantum Monte Carlo program

for electronic structure. *Journal of Computational Physics* 228(9):3390-3404.
32. Dovesi R, *et al.* (2014) CRYSTAL14: A program for the ab initio investigation of crystalline solids. *International Journal of Quantum Chemistry* 114(19):1287-1317.
33. Burkatzki M, Filippi C, & Dolg M (2007) Energy-consistent pseudopotentials for quantum Monte Carlo calculations. *The Journal of Chemical Physics* 126(23):234105.
34. Umrigar CJ, Wilson KG, & Wilkins JW (1988) Optimized trial wave functions for quantum Monte Carlo calculations. *Physical Review Letters* 60(17):1719-1722.
35. Umrigar CJ, Nightingale MP, & Runge KJ (1993) A diffusion Monte Carlo algorithm with very small time‐step errors. *The Journal of Chemical Physics* 99(4):2865-2890.
36. Zen A, Sorella S, Gillan MJ, Michaelides A, & Alfè D (2016) Boosting the accuracy and speed of quantum Monte Carlo: Size consistency and time step. *Physical Review B* 93(24):241118.
37. Purwanto W, Krakauer H, & Zhang S (2009) Pressure-induced diamond to $\ensuremath{\beta}$-tin transition in bulk silicon: A quantum Monte Carlo study. *Physical Review B* 80(21):214116.
38. Spanu L, Sorella S, & Galli G (2009) Nature and Strength of Interlayer Binding in Graphite. *Physical Review Letters* 103(19):196401.
39. Ceperley DM & Alder BJ (1980) Ground State of the Electron Gas by a Stochastic Method. *Physical Review Letters* 45(7):566-569.
40. Adamo C & Barone V (1999) Toward reliable density functional methods without adjustable parameters: The PBE0 model. *The Journal of Chemical Physics* 110(13):6158-6170.
41. Yang K, Zheng J, Zhao Y, & Truhlar DG (2010) Tests of the RPBE, revPBE, τ-HCTHhyb, ωB97X-D, and MOHLYP density functional approximations and 29 others against representative databases for diverse bond energies and barrier heights in catalysis. *The Journal of Chemical Physics* 132(16):164117.
42. del Campo JM, Gázquez JL, Trickey SB, & Vela A (2012) Non-empirical improvement of PBE and its hybrid PBE0 for general description of molecular properties. *The Journal of Chemical Physics* 136(10):104108.
43. Hu T & Dong J (2015) Structural phase transitions of phosphorene induced by applied strains. *Physical Review B* 92(6):064114.
44. Signorello G, Karg S, Björk MT, Gotsmann B, & Riel H (2013) Tuning the Light Emission from GaAs Nanowires over 290 meV with Uniaxial Strain. *Nano Letters* 13(3):917-924.
45. Schulten K, Schulten Z, & Szabo A (1981) Dynamics of reactions involving diffusive barrier crossing. *The Journal of Chemical Physics* 74(8):4426-4432.